\documentclass[pra,a4paper,twocolumn,showpacs,superscriptaddress]{revtex4}
\usepackage{amsmath}
\usepackage{amsfonts}
\usepackage{graphicx}
\usepackage{longtable}
\newcommand{\be}{\begin{equation}}
\newcommand{\ee}{\end{equation}}

\newcommand{\ba}[1]{\left(\begin{array}{#1}}
\newcommand{\ea}{\end{array}\right)}
\begin{document}

\title{From the quantum relative Tsallis entropy to its conditional form: Separability criterion beyond local and global spectra} 
\author{A. K. Rajagopa1}\affiliation{Inspire Institute Inc., Alexandria, Virginia, 22303, USA.}
\affiliation{Harish-Chandra Research Institute, Chhatnag Road, Jhunsi, Allahabad 211 019, India}
\author{Sudha }
\affiliation{Department of Physics, Kuvempu University, 
Shankaraghatta, Shimoga-577 451, India}
\affiliation{Inspire Institute Inc., Alexandria, Virginia, 22303, USA.}
\author{ Anantha S Nayak}
\affiliation{Department of Physics, Kuvempu University, 
Shankaraghatta, Shimoga-577 451, India}
\author{A. R. Usha Devi}
\affiliation{ Department of Physics, Bangalore University, Bangalore 560 056, India}
\affiliation{Inspire Institute Inc., Alexandria, Virginia, 22303, USA.}
\date{\today}
\begin{abstract}
The quantum relative R\'{e}nyi entropy of two non-commuting density matrices was recently defined from  which its conditional entropy is deduced. This framework is here extended to the corresponding Tsallis relative entropy and to its conditional form. This new expression of Tsallis
conditional entropy is shown to witness entanglement beyond the method based on global and local
spectra of composite quantum states. When the reduced density matrix happens to be a maximally mixed state, this conditional entropy coincides with the expression in terms of Tsallis entropies derived earlier by Abe and Rajagopal (Physica A {\bf 289}, 157 (2001)). Separability range in noisy one parameter family of W and GHZ states with $3$ and $4$ qubits is explored here and it is  shown that the results inferred from negative Tsallis conditional entropy matches with that obtained through Peres' partial transpose criteria for one-parameter family of W states, in one of its partitions.   The criteria is shown to be non-spectral through its usefulness in identifying entanglement in isospectral density matrices.    
\end{abstract}
\maketitle
\section{Introduction} 
Entropic characterization of separability of composite quantum systems has attracted significant attention~\cite{ent,hki,aber,tsa,sabe,canosa,jb,prabhu,arss}. This is based on the observation that von Neumann conditional entropy of a pure bipartite entangled state is negative and this points towards a more general feature that entangled systems could be more disordered locally than globally while it is not possible for separable states~\cite{nk}. Positivity of generalized quantum conditional entropies such as R\'{e}nyi and Tsallis conditional entropies is a more effective tool to investigate global vs local disorder in mixed states and it  leads to  stricter constraints on separability than that obtained using conditional von Neumann entropy~\cite{aber,tsa,sabe,canosa,jb,prabhu}. However, entropic criterion is not sufficient for separability as it has been shown that any criterion based on global and local spectra is incapable of distinguishing a separable state from entangled ones~\cite{nk}.          

More recently a new quantum generalization of  R\'{e}nyi relative entropy has been introduced~\cite{mw,mds} and there is a surge of activity in establishing several properties of this new version of R\'{e}nyi entropy~\cite{mw,mds,fl,sb,aer}. 
This generalized R\'{e}nyi relative entropy of a pair of density  operators reduces to the traditional one when the two density operators commute with each other. It is thus natural to anticipate that this quantity is more effective than its traditional version when noncommuting density matrices are involved. In Ref.\cite{mw}, this generalization was termed ``sandwiched'' R\'{e}nyi relative entropy in view of the sandwiching of non-commutative operators in a particular way and hence this nomenclature is an appropriate one for the generalized R\'{e}nyi relative entropy.  

In this paper we introduce sandwiched version of Tsallis relative entropy and identify a new quantum version of conditional Tsallis entropy derived therefrom. We investigate entanglement based on non-positive values of sandwiched Tsallis conditional entropy  defined here. An earlier version of Tsallis conditional entropy,  introduced by Abe and Rajagopal (AR)~\cite{aber} using classical conditional Tsallis probabilities,  has been found to be useful in exploring separability of several families of composite quantum systems~\cite{aber,tsa,sabe,canosa,jb,prabhu}. The conditional version of the sandwiched  Tsallis relative entropy (CSTRE) offers as a generalized version when the density matrix and its reductions do not commute with each other. The new version reduces to the AR q-conditional entropy when the reduced density matrix is  maximally disordered. We show that the identification of inseparability based on non-positive values of CSTRE goes beyond the  spectral criterion and it is capable of distinguishing separable states from  entangled states that are globally and locally isospectral.

The contents of the paper are organized into five sections, including the introduction and motivation behind this work detailed here in Section~I. In Section II we introduce sandwiched relative Tsallis entropy  and its conditional version. We employ this conditional form to examine the separability range in one-parameter families of  W and GHZ states in Section III. In Section IV we show that the CSTRE approach of identifying separability  goes beyond spectral disorder criteria and can distinguish separable states from entangled ones when they share same global and local spectra. Section V contains summary and concluding remarks.
\section{Sandwiched Relative Tsallis Entropy and its conditional version} 
The generalized entropies, the R\'{e}nyi and Tsallis entropies, denoted  respectively by $S^{R}_q(\rho)$, $S^{T}_q(\rho)$ are given by~\cite{hki, aber, tsa1}  
\begin{eqnarray}
\label{rt} 
S^{R}_q(\rho)&=& \frac{1}{1-q} \log \mbox{Tr}[\rho^q] \nonumber \\
S^{T}_q(\rho)&=& \frac{\mbox{Tr}[\rho^q]-1}{1-q}.   
\end{eqnarray}
where $q$ is a real positive parameter.  Both these reduce to von-Neumann entropy in the limit $q\rightarrow 1$.   

The traditional quantum relative R\'{e}nyi entropy for a pair of density operators $\rho$ and $\sigma$ is defined, by ignoring the ordering of the density matrices, as
\begin{eqnarray}
\label{rre}
D_q^{R}(\rho\vert\vert\sigma)&=&\frac{\log \mbox{Tr}\left( \rho^q \sigma^{1-q} \right)}{q-1} \ \ \mbox{if}\ q\in (0,1) \cup (1, \infty)  \\
&=& \mbox{Tr}\left[\rho(\log \rho-\log \sigma) \right]\ \ \mbox{when}\ \ q\rightarrow 1; \nonumber
\end{eqnarray}

Recently a generalized version of quantum relative R\'{e}nyi entropy was introduced by Wilde et al.~\cite{mw} and M\"{u}ller-Lennert et al.~\cite{mds} independently:       
\begin{eqnarray}
\label{srre}
\tilde{D}_q^{R}(\rho\vert\vert\sigma)&=&\frac{1}{q-1}\log\left[ \mbox{Tr}\left(\sigma^{\frac{1-q}{2q}} \rho \sigma^{\frac{1-q}{2q}}\right)^q\right] \\
& & \mbox{where}\  q\in (0,1)\cup (1,\infty). \nonumber 
\end{eqnarray} 
The quantum relative R\'{e}nyi entropy (\ref{srre}) reduces to the traditional one given by (\ref{rre}) when the density matrices $\rho$ and $\sigma$ commute and hence the new version is an extension to non-commutative case.  

We now consider the traditional form of Tsallis relative entropy
\be
D_q^{T}(\rho\vert\vert\sigma)=\frac{\mbox{Tr}\left( \rho^q \sigma^{1-q} \right)-1}{1-q},
\ee
and define the `sandwiched' Tsallis relative entropy in a similar manner,    
\be
\label{gstre}
\tilde{D}_q^{T}(\rho\vert\vert \sigma)=\frac{\mbox{Tr}\left\{\left(\sigma^{\frac{1-q}{2q}}\rho \, \sigma^{\frac{1-q}{2q}}\right)^q \right\}-1}{1-q}
\ee

Note that when $\sigma=I$, sandwiched Tsallis relative entropy (\ref{gstre}) reduces to the Tsallis entropy $S^{T}_q(\rho)=\frac{\mbox{Tr}(\rho^q)-1}{1-q}$ and in the limit $q\rightarrow 1$, it reduces to the  von-Neumann relative entropy: $\lim_{q\rightarrow 1}\, D_q(\rho||\sigma)=\mbox{Tr}(\rho(\log\rho-\log\sigma))$\cite{note1}.

We now define the conditional version of $\tilde{D}_q^{T}(\rho_{AB}\vert\vert \sigma)$ by taking $\sigma \equiv I_A\otimes \rho_B$ (or $\rho_A\otimes I_B$) where $\rho_B={\rm Tr}_A[\rho_{AB}]$  ($\rho_A={\rm Tr}_B[\rho_{AB}]$) is the subsystem density matrix of  the composite state $\rho_{AB}$. It is given by   
\be
\label{cstre}
\tilde{D}^{T}_q(\rho_{AB}||\rho_B)=\frac{\tilde{Q}_q(\rho_{AB}||\rho_B)-1}{1-q}
\ee
where 
\be
\label{qts}
\tilde{Q}_q(\rho_{AB}||\rho_B)=\mbox{Tr}\left\{\left((I\otimes\rho_B)^{\frac{1-q}{2q}}\rho_{AB}(I\otimes\rho_B)^{\frac{1-q}{2q}}\right)^q\right\}.
\ee

The sandwiched conditional Tsallis  entropy (\ref{cstre}) reduces to AR q-conditional Tsallis entropy~\cite{aber} 
\be 
\label{ar}
 S_q^{T}(A\vert B)=\frac{1}{q-1}\left(1-\frac{\mbox{Tr}\rho_{AB}^q}{\mbox{Tr}\rho_B^q} \right)
\ee
when the subsystem density matrix is a maximally mixed state.  Negative value of AR q-conditional entropy indicate entanglement and it has been employed as a separability criterion for several classes of composite states~\cite{aber,tsa,sabe,canosa,jb,prabhu}. Here we employ the new form of sandwiched conditional Tsallis entropy derived from its relative entropy to infer about entanglement.    

While the evaluation of the expression $\tilde{Q}_q(\rho_{AB}||\rho_B)$ does not seem trivial, construction of the unitary operator that diagonalizes the subsystem density matrix $\rho_B$ makes the calculation a feasible one. If $U_B$ is the unitary operator that diagonalizes $\rho_{B}$, we have      
\begin{eqnarray}
\sigma_D&=&U_\sigma\,\sigma^{\frac{1-q}{2q}}U_\sigma^\dag \ \   \mbox{where}  \\ 
\sigma&=& I\otimes \,\rho_B, \ U_\sigma=I\otimes U_B \ \mbox{and} \nonumber \\  
\sigma_D&=& \mbox{diag}\left(\lambda_1^{\frac{1-q}{2q}}\cdots\lambda_n^{\frac{1-q}{2q}}\right). \nonumber 
\end{eqnarray}
Thus the expression for $\tilde{Q}_q(\rho_{AB}||\rho_B)$ in Eq. (\ref{qts}) simplifies to
\begin{eqnarray}
\tilde{Q}_q(\rho_{AB}||\rho_B)&=&\mbox{Tr}\left\{\left( \sigma_D U_\sigma \rho U_\sigma^\dag \sigma_D \right)^q \right\}
\end{eqnarray}
Denoting  $\Gamma=\sigma_D U_\sigma \rho U_\sigma^\dag \sigma_D$, an evaluation of the eigenvalues $\gamma_i$ of $\Gamma$ immediately leads us to the quantity $\tilde{Q}_q(\rho_{AB}||\rho_B)$ as $\sum_{i=1}^d \, \gamma_i^q$, $d$ being the dimension of $\rho$. Thus we finally obtain an expression for the conditional form of sandwiched Tsallis relative entropy (CSTRE);
\be 
\label{f}
\tilde{D}^{T}_q (\rho_{AB}||\rho_B)=\frac{\sum_{i=1}^d \, \gamma_i^q-1}{1-q}. 
\ee
We make use of Eq. (\ref{f}) to determine the separability range in one-parameter families of W and GHZ states in the following section.  

While the relation $\tilde{D}_q^{T}(\rho\vert\vert \sigma)\leq   D_q^{T}(\rho\vert\vert \sigma)$ for all $q\geq 1$ follows from  Lieb-Thirring inequality~\cite{mw}, $D_q^{T}(\rho\vert\vert \sigma)$ leads to AR-like q-conditional entropy when $\rho$, $\sigma$ commute. The inequality $\tilde{D}_q^{T}(\rho\vert\vert \sigma)\leq   D_q^{T}(\rho\vert\vert \sigma)$ gives the hint that the separability criterion deduced from $\tilde{D}_q^{T}(\rho\vert\vert \sigma)$ is stricter than or equal to that obtained from $D_q^{T}(\rho\vert\vert \sigma)$.  
In view of the fact that AR criterion is weaker (or identical to) the PPT criterion, we may conjecture that there is a nesting of criteria of the form \\
AR-like criterion $\geq$ CSTRE criterion $\geq$ PPT. 
\section{Separability of noisy one-parameter families of W and GHZ states} 
The symmetric one parameter family of $N$-qubit mixed states, involving a W- or GHZ- state are respectively given by 
\be
\label{w}
\rho_N^{(W)}(x)=\left(\frac{1-x}{N+1}\right)P_N+ x\vert W \rangle_N \langle W \vert
\ee
and 
\be
\label{ghz}
\rho_N^{(GHZ)}(x)=\left(\frac{1-x}{N+1}\right)P_N+ x\vert GHZ \rangle_N \langle GHZ \vert.
\ee
Here $0\leq x \leq 1$ and $P_N=\sum_M \, \vert N/2,\,M \rangle \langle N/2,\,M \vert$ denotes the projector onto the symmetric subspace of $N$-qubits spanned  by the $N+1$ angular momentum states $\vert N/2,\,M \rangle$, $M=N/2,N/2-1,\cdots ,-N/2$ belonging to the maximum value $J=N/2$ of total angular momentum.  

A systematic attempt to examine the separability range of the noisy one parameter family of W and GHZ states using the AR q-conditional entropy has been carried out in Ref.~\cite{prabhu}. While they could obtain a result matching with that of positive partial transpose (PPT) criterion~\cite{peres} for 
$2$-qubit states of $\rho_{N=2}^{(W)}(x)$ the range of separability identified by them is weaker than that through PPT criterion, for both W, GHZ families when $N\geq 3$. Here we identify that non-commutativity of the density matrix $\rho_{AB}$ with its subsystem state $\rho_B$ does indeed play a role and the separability domain inferred through non-negative values of CSTRE is stricter compared to that obtained from AR q-conditional entropy, though it is  weaker than the PPT criterion in some cases.  

For $N=3$, a direct evaluation of the subsystem density matrix $\rho_{BC}$ of $\rho^{W}_{3}(x)\equiv \rho_{ABC}$ and the unitary matrix that diagonalizes it leads us to $\Gamma=\sigma_D U_\sigma \rho U_\sigma^\dag \sigma_D$. Here, we have, 
\begin{eqnarray}
\sigma_D&=&I_2\otimes \mbox{diag}\left(  \left(\frac{1}{3}\right)^{\frac{1-q}{2q}},0,\left(\frac{1-x}{3}\right)^{\frac{1-q}{2q}},\left(\frac{1+x}{3}\right)^{\frac{1-q}{2q}} \right) \nonumber \\
U_\sigma&=&I_2\otimes \ba{cccc} 1 & 0 & 0 & 0 \\ 0 & \frac{1}{\sqrt{2}} & -\frac{1}{\sqrt{2}} & 0 \\ 0 & 0 & 0 & 1 \\  0 & \frac{1}{\sqrt{2}} & \frac{1}{\sqrt{2}} & 0 \ea \  \mbox{and} \  \rho=\rho^{W}_{3}(x). \nonumber
\end{eqnarray}
The non-zero eigenvalues of $\Gamma$ are found to be,  
\begin{eqnarray}
\gamma_1&=&\frac{3(1-x)3^{-\frac{1}{q}}}{4},\ \gamma_2=\frac{3(1-x)^{\frac{1}{q}}3^{-\frac{1}{q}}}{4} \nonumber \\
\gamma_3&=&\frac{3^{-\frac{1}{q}}(1+3x)\left(1+x+2(1+x)^{\frac{1}{q}}\right)}{4(1+x)}  \\ 
\gamma_4&=& \frac{3^{-\frac{1}{q}}\left((1+x)(1-x)^{\frac{1}{q}}+2(1-x)(1+x)^{\frac{1}{q}}\right)}{4(1+x)} \nonumber
\end{eqnarray}     
One can now readily evaluate the expression for CSTRE in Eq.(\ref{f}) and for different values of $q$, we obtain $\tilde{D}^{T}_q(\rho^{(W)}_3(x)\vert\vert \rho_{BC})$ as a function of $x$. The following plots (Figs. 1 and 2) illustrate the stricter separability range for $\rho^{(W)}_3(x)$, in its A:BC partition, for increasing values of $q$. 
\begin{figure}[ht]
\includegraphics* [width=2.4in,keepaspectratio]{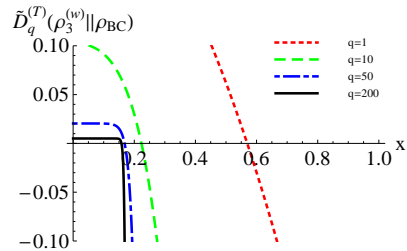} 
\caption{(Color Online)The conditional form of sandwiched Tsallis relative entropy $\tilde{D}^{T}_q(\rho^{(W)}_3(x)\vert\vert \rho_{BC})$ for one-parameter family of three qubit W-states  as function of $x$ for different values of $q$. It can be seen that CSTRE is negative for $x\geq 0.57$ when $q=1$, whereas it is negative as  $x\rightarrow 0.1547$  (the value of $x$ identified by PPT criterion) for $q>>1$. 
All the quantities are dimensionless.}
\end{figure}
\begin{figure}[ht]
\includegraphics* [width=2.4in,keepaspectratio]{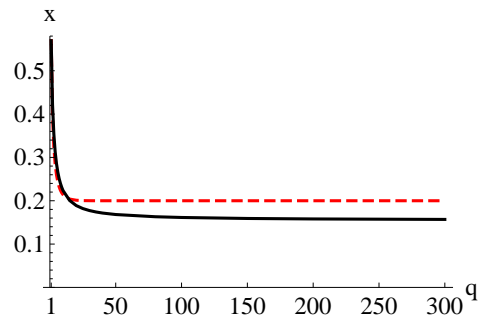} 
\caption{(Color Online) Implicit plot of $\tilde{D}^{T}_q(\rho^{(W)}_3\vert\vert \rho_{BC})=0$ as a function of $q$ (solid line) indicating that  
$x\rightarrow 0.1547$ as $q\rightarrow\infty$. In contrast, the implicit plot of Abe-Rajagopal q-conditional enltropy $S_q^{T}(A\vert BC)=0$ (dashed line) leads to $x\rightarrow 0.2$ as $q\rightarrow\infty$. The quantities plotted are dimensionless.}  
\end{figure}
It can be readily seen through Figs. 1 and 2 that the separability range $0\leq x \leq 0.1547$ in the A:BC partition of the one-parameter family of $3$-qubit W states obtained through CSTRE approach, is in complete agreement with that obtained through partial transpose criterion. It is to be noticed (See Fig. 2) that AR q-conditional entropy yields a weaker separability range~\cite{prabhu} $0\leq x \leq 0.2$  for the A:BC partition of  $\rho_3^{(W)}(x)$. 

In a similar manner we evaluate the CSTRE $\tilde{D}^{T}_q(\rho^{(W)}_4(x)\vert\vert \rho_{BCD})$ and arrive at the separability range in the A:BCD partition of the state $\rho_4^{(W)}(x)$. It is seen that $\rho_4^{(W)}(x)$ is separable when $x\leq 0.1124$ in complete conformity with the separability range obtained through PPT criterion. In Fig.~3 we have illustrated our result  for $\rho_4^{(W)}(x)$ through an implicit plot of $\tilde{D}^{T}_q(\rho^{(W)}_4\vert\vert \rho_{BCD})=0$ and compare it with that of AR q-conditional entropy $S_q^{T}(A\vert BCD)=0$. 
\begin{figure}[ht]
\includegraphics* [width=2.4in,keepaspectratio]{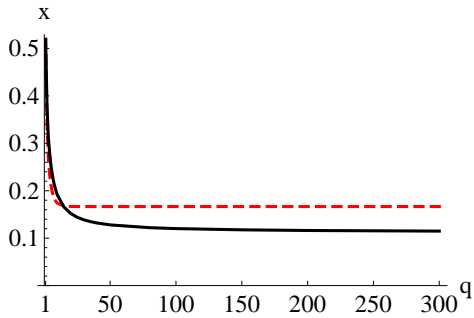} 
\caption{(Color Online) The implicit plots of $\tilde{D}^{T}_q(\rho^{(W)}_4\vert\vert \rho_{BCD})=0$ (solid line) and $S_q^{T}(A\vert BCD)=0$ (dashed line) as a function of $q$. Here $x\rightarrow 0.1124$ according to the CSTRE appoach while $x\rightarrow 0.1666$ inferred from AR q-conditional entropy,  both in the limit $q\rightarrow\infty$. The quantities plotted are dimensionless.}  
\end{figure}

It is pertinent to point out that the state $\rho^{(W)}_N(x)$ and its reduced counterparts $I_2\otimes \rho_{N-1}$ are non-commuting thus offering themselves as ideal examples to test the new CSTRE criterion for separability. 

The separability range of one parameter family of $N$-qubit GHZ states has been examined using AR q-conditional entropy in Ref.~\cite{prabhu} and they had obtained the separability range matching with that from PPT criteria for $\rho^{(GHZ)}_3(x)$ only in the A:BC partition. It may be noted that  
the separability range in the A:BC partition of $\rho_3^{(GHZ)}(x)$ is $[0,\frac{1}{7}]$ through PPT as well as AR q-conditional entropy criteria. An explicit evaluation using the CSTRE approach is seen to reproduce the same separability range indicating that the CSTRE separability domain is bounded by the PPT range.  

For $N=4$ also, the separability ranges obtained through both PPT criterion and AR q-conditional entropy approach match with each other only in the A:BCD partition of $\rho_4^{(GHZ)}(x)$. Here, we identify the separability ranges through the new CSTRE approach and show that a separability range same as that through AR q-conditional approach is obtained  in all possible partitions of the state $\rho^{(GHZ)}_4(x)$. Table~1 summarizes our results for the different partitions of the one parameter family of W-, GHZ- states. 
 \begin{table}[ht]
\caption{Comparison of separability range of one parameter families of states through entropic criteria and PPT}
\scriptsize{
\begin{tabular}{|c|c|c|c|c|}
\hline
Quantum  & Von-Neumann & AR  &  &  \\ 
State & conditional  & q-conditional & CSTRE & PPT \\
 & entropy & entropy & & \\
\hline\hline 
$\rho^{(W)}_{3}$ &    &  &   &  \\ 
\cline{1-1}A:BC partition  &  $\{ 0,\, 0.5695\}$  & $\{0,\, 0.2\}$ & $\{0,\, 0.1547\}$  & $\{0,\, 0.1547\}$ \\ \hline 
AB:C partition  &  $\{0,\, 0.7645\}$  & $\{0,\, 0.4286\}$ & $\{0,\, 0.3509\}$  & $\{0,\, 0.1547\}$ \\ \hline  
$\rho^{(GHZ)}_{3}$ & & & & \\ 
\hline
\cline{1-1}A:BC partition  &  $\{ 0,\, 0.5482\}$  & $\{0,\, 1/7\}$ & $\{0,\,1/7\}$  & $\{0,\,1/7\}$ \\ \hline 
AB:C partition &  $\{0,\, 0.7476\}$  & $\{0,\, 1/3\}$ & $\{0,\,1/3\}$  & $\{0,\, 1/7\}$ \\ \hline 
$\rho^{(W)}_{4}$ &    &  &   &  \\  
\cline{1-1} A:BCD partition  &  $\{0,\, 0.5193\}$  & $\{0.\,0.1666\}$ & $\{0,\,0.1123\}$  & $\{0,\, 0.1123\}$ \\ \hline 
AB:CD partition  & $\{ 0,\, 0.6560\}$   &  $\{0,\, 0.2105\}$ & $\{0,\,0.2105\}$ &  $\{0,\, 0.0808\}$       \\ \hline 
ABC:D partition  &  $\{0,\, 0.8222\}$  & $\{0,\, 0.5454\}$  & $\{0,\, 0.4174\}$  & $\{0,\, 0.1123\}$ \\ \hline 
$\rho^{(GHZ)}_{4}$ &    &  &  &  \\ 
\cline{1-1}A:BCD partition  &  $\{0,\, 0.4676\}$  & $\{0,\, 0.0909\}$ & $\{0,\, 0.0909\}$  & $\{0,\,  0.0909\}$ \\ \hline 
AB:CD partition  &  $\{0,\, 0.6560\}$  & $\{0,\, 0.2105\}$ & $\{0,\, 0.2105\}$  & $\{0,\, 0.0625\}$ \\ \hline 
ABC:D partition  &  $\{0,\, 0.7868\}$  & $\{0,\, 0.375\}$ & $\{0,\, 0.375\}$  & $\{0,\, 0.0909\}$ \\ \hline 
\end{tabular}}
\end{table} 

It can be readily seen through Table~1 that the CSTRE approach yields a separability range that is either equal to or {\em {stricter}} than the range obtained through AR q-conditional entropy and it matches with PPT criterion in some of the non-commuting cases such as in $\rho^{(W)}_{N=3,\,4}$ in one of their (A:BC, A:BCD) partitions.  

We proceed further in Sec.~IV to illustrate that  CSTRE separabilty criterion is non-spectral in nature and thus can distinguish separable and entangled states which  share same local and global eigenvalues.

\section{Non-spectral nature of CSTRE criterion}  

Generalized  entropies serve as a measure of disorder in a given quantum state and negative values of traditional versions of generalized conditional entropies point towards more global order than local order in a composite system. Separable states are more locally ordered than globally as the eigen spectra of the whole composite separable state is majorized by that of its reduced systems~\cite{nk}. Negative values of conditional entropies reflect the contrasting feature that local spectra {\em need not} majorize global spectra in entangled states. However, separabilitly criterion based purely on the spectra of composite and reduced states is shown to be insufficient~\cite{nk}.  This feature was illustrated through an  example of an isospectral pair of two qubit states $\rho_{AB}$ and $\varrho_{AB}$, which  share same local and global spectra with $\rho_{AB}$ being entangled while $\varrho_{AB}$ is separable~\cite{nk}: 
\begin{eqnarray} 
\label{isoe1}
\rho_{AB}&=&\frac{1}{3}\ba{cccc} 1 & 0 & 0 & 0 \\ 
                               0 & 1 & 1 & 0  \\ 
                               0 & 1 & 1 & 0  \\
                               0 & 0 & 0 & 0 \\ \ea  \\ 
\label{isoe2}                               
\varrho_{AB}&=&\frac{1}{3}\ba{cccc} 1 & 0 & 0 & 0 \\ 
                               0 & 0 & 0 & 0  \\ 
                               0 & 0 & 0 & 0  \\
                               0 & 0 & 0 & 2 \\ \ea. 
\end{eqnarray}
It is worth observing that  the separable state $\varrho_{AB}$ commutes with its reduced density matrices  whereas the entangled state $\rho_{AB}$ and its subsystems are non-commutative. And interestingly,  the new sandwiched conditional Tsallis entropy is capable of distinguishing the isospectral states and hence the approach proves to be superior to any spectral disorder criteria. We demonstrate the effectiveness of CSTRE separability criterion in the above examples of two qubit isospectral states.  

For the state $\rho_{AB}$ we obtain, 
\begin{eqnarray*}
\gamma_1&=&2^{\frac{1-q}{q}}3^{-\frac{1}{q}} \nonumber \\
 \gamma_2&=&\left(1+2^{\frac{1-q}{q}}\right)3^{-\frac{1}{q}}  
\end{eqnarray*} 
and  CSTRE is given by 
\be
\label{isoe3}
\tilde{D}^{T}_q (\rho_{AB}||\rho_B)=\frac{\left(1+2^{\frac{1-q}{q}}\right)^q+2^{1-q}-3}{3(1-q)}.  
\ee
A plot of $\tilde{D}^{T}_q (\rho_{AB}||\rho_B)$ as a function of $q$ is given in Fig.~4. 
\begin{figure}[ht]
\includegraphics* [width=2.4in,keepaspectratio]{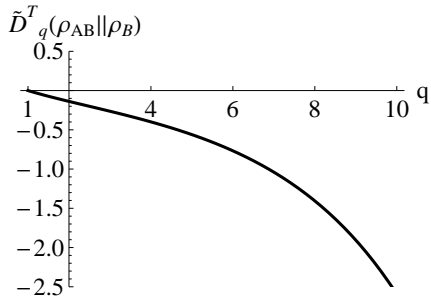} 
\caption{A plot of CSTRE $\tilde{D}^{T}_q (\rho_{AB}||\rho_B)$ of the two qubit state (\ref{isoe1}) as a function of $q$. It is readily seen that CSTRE is negative for all values of $q>1$ indicating that the state is entangled. Both the quantities are dimensionless.}
\end{figure} 
It can be readily seen through Fig. 4 that  $\tilde{D}^{T}_q (\rho_{AB}||\rho_B)$ is negative for all values of  $q$, except for $q=1$

On the otherhand,  in the two qubit separable state $\varrho_{AB}$ we find that   
\begin{eqnarray}
\gamma_1&=&\left(\frac{2}{3}\right)^{\frac{1}{q}}, \ \ \gamma_2=\left(\frac{1}{3}\right)^{\frac{1}{q}}
\end{eqnarray}
 leading to $\gamma_1^q+\gamma_2^q=1$ and hence, 
 \be
 \tilde{D}^{T}_q (\varrho_{AB}||\varrho_B)=0 
 \ee 
for all values of $q$. 
 
The isospectral  example of two qubit states highlight that CSTRE approach is essentially non-spectral in nature, unlike other entropic criteria~\cite{note2}. 

\section{Conclusion} 
Characterization of  entanglement based on R\'enyi and Tsallis conditional entropies~\cite{ent,hki,aber,tsa,sabe,canosa,jb,prabhu,arss} is essentially based on the spectra of the composite state and its subsystems. Separable states are more disordered locally than globally~\cite{nk} and this feature is reflected through  their generalized  conditional entropies being positive. Negative values of conditional entropies imply entanglement. However, spectral criteria is only  sufficient but not necessary to detect entanglement. There exist examples of separable and entangled states which share same global and local spectra~\cite{nk}.  

Motivated by the recently introduced sandwiched R\'enyi relative entropy~\cite{mw, mds}, we defined corresponding version of Quantum Tsallis relative entropy for a pair of non-commuting density matrices.  We have shown that conditional Tsallis entropy  derived from sandwiched relative entropy of a composite quantum state and its subsystem is useful to characterize entanglement beyond the spectral disorder criteria. The new CSTRE reduces to the traditional form of Tsallis conditional entropy (AR q-conditional entropy)  developed by Abe and Rajagopal~\cite{aber} when the subsystem density matrix is a maximally mixed state. We have used CSTRE to investigate separability of noisy one parameter families of $3$ and $4$ qubit W, GHZ states. The results are compared with those obtained by AR q-conditional entropy and also Peres' PPT criterion. It is identified that CSTRE is superior to AR q-conditional entropy, while the separability range is limited by that drawn from PPT criterion. These results are collected together in Table~I. The CSTRE approach is shown here to be useful to distinguish between  isospectral separable and entangled states. While the CSTRE approach is seen to be either identical or weaker than PPT criterion in the examples considered here, it is an open question whether this hierarchy is true for all states and whether this appoach can identify bound entangled states.      
\section*{Acknowledgement:}  
Anantha S. Nayak acknowledges the support of Department of Science and Technology (DST), Govt. of India through the award of INSPIRE fellowship. \\ 

This work is dedicated to Professor Constantino Tsallis on the occasion of his 70th Birthday.  
 

\begin{thebibliography}{88}
\bibitem{note1} It may be noted that when $q=1/2$ Eq. (\ref{gstre}) reduces to $\tilde{D}_{\frac{1}{2}}^{T}(\rho||\sigma)=2(F(\rho|\sigma)-1)$ which is related to the Bures metric.  
Here $F(\rho|\sigma)=\mbox{Tr} \left(( B^\dagger B\right)^{\frac{1}{2}}$, $B=\rho^{1/2}\sigma^{1/2}$ is the Uhlmann fidelity function written in the form of a norm. More generally, from Eq. (\ref{gstre}), we observe that $\tilde{D}_{q}^{T}(\rho||\sigma)$ can also be expressed as a q-norm $\tilde{D}_{q}^{T}(\rho||\sigma)=\frac{\mbox{Tr} \left(X^\dagger X\right)^{q}-1}{1-q}$ with $X=\rho^{\frac{1}{2}}\sigma^{\frac{1-q}{2q}}$ and may be related to a more general form of $q$-Bures metric, in analogy with the case where $q=1/2$ as noted above reduces to the usual Bures metric.
\bibitem{note2} Note  that AR q-conditional entropy of the isospectral states $\rho_{AB}$, $\varrho_{AB}$ is zero for all values of $q$ and so, characterizing  separability based on AR q-conditional entropy was not useful.
\end{thebibliography}
\end{document}